\documentstyle[12pt,epsfig]{article}

\newcommand{\be}{\begin{equation}}
\newcommand{\ee}{\end{equation}}

 1
\font\elevenrm=cmr10 scaled\magstep 1
 1

\def\reff{\hang\noindent}

\begin{document}

\def\pdot {\dot P}
\def\Omdot {\dot \Omega}
\def\ltsima{$\; \buildrel < \over \sim \;$}
\def\lsim{\lower.5ex\hbox{\ltsima}}
\def\gtsima{$\; \buildrel > \over \sim \;$}
\def\gsim{\lower.5ex\hbox{\gtsima}}
\def\msole{~M_{\odot}}
\def\mdot {\dot M}
\def\uu {4U~0142$+$61~}
\def\oo {1E~1048.1$-$5937~}
\def\kes {1E~1841$-$045~}
\def\axj {AX~J1845.0$-$0300~}
\def\rx {1RXS~J170849$-$400910~}
\def\ee {1E~2259$+$586~}
\def\saxj  {SAX J1808.4$-$3658~}

\vspace*{.8cm}

  \centerline{\bf THE ZOO OF X-RAY PULSARS}

\vspace{1cm}
  \centerline{ Sandro Mereghetti}
\vspace{0.4cm}
  \centerline{Istituto di Fisica Cosmica G.Occhialini -- CNR  }
  \centerline{\elevenrm via Bassini 15, Milano, ITALY }
\vspace{1.cm}  
  \centerline{\it Invited Talk presented at the Workshop }
   \centerline{\it   ``Frontier Objects in Astrophysics
and Particle Physics''}  
 \centerline{\it    Vulcano (Italy), May 22-27, 2000 }

\vspace{1.cm}
\begin{abstract}
I review some recent developments in the field of    X--ray pulsars:
the discovery of millisecond pulsations in the Low Mass Binary System
SAX J1808.4-3658, the large number of transient Be systems discovered in
the Magellanic Clouds and the enigmatic class of objects known as Anomalous
X--ray Pulsars.
\end{abstract}
\vspace{2.0cm}

\section{ Introduction }

Accretion powered X--ray pulsars   were among the first sources observed in X--ray
astronomy and, thanks to their characteristic timing signatures,
could be quite soon correctly interpreted as rotating,
magnetized neutron stars
(Pringle \& Rees 1972, Davidson \& Ostriker 1973). Since then, the observation
and study of X--ray pulsars has provided a wealth of important
information on the physics of neutron stars and on the evolution
of stars in binary systems.

The designation of \textit{``X-ray'' pulsars} has traditionally been used
to indicate the objects powered by accretion of matter from a
companion star in a close, interacting binary (such as, e.g., Her
X-1 and Cen X-3), in contrast to  the \textit{``radio'' pulsars},
consisting of (in general) isolated neutron stars, the emission
of which is powered by the loss of rotational energy. In fact
only a few of the youngest and more powerful (in terms of E$_{rot}$)
radio pulsars were observed at X-ray energies (e.g. the Crab and
Vela pulsars). Nowadays it appears more appropriate to
distinguish between \emph{accretion powered} and \emph{rotation
powered} pulsars, since the better sensitivity of current
satellites has allowed the detection of X-rays from $\sim$40 of
the more than 1000 radio pulsars.

Here we concentrate only on accretion powered pulsars (see
Becker 2000 for a review of the X-ray properties of rotation
powered pulsars).

\begin{figure}
\centerline{\psfig{figure=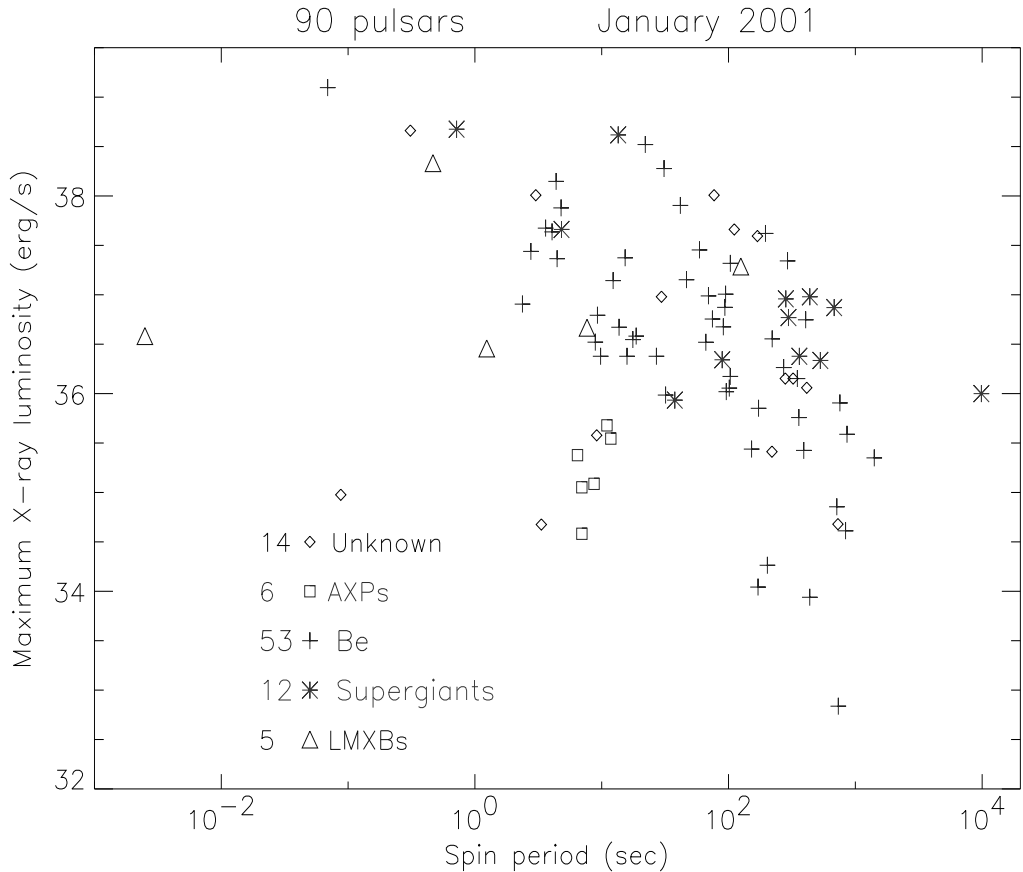 ,width=15cm,height=12cm,angle=90} }
\caption{Spin period and maximum X--ray luminosity of   accretion
powered X--ray pulsars. The different symbols indicate the various classes
of pulsars.}
\label{pvslx}
\end{figure}

The current census of accretion powered X--ray pulsars lists
$\sim$95 objects.  28 of them  are located in the Magellanic
Clouds (see Table 1) and three possible pulsars have
also been reported in more distant galaxies
(M31 and M33, Israel et al. 1995, Dubus et al. 1999). The observed spin
periods are in the range from 2.5 ms to about 3 hours, but most of them are
between $\sim$1 and 1000 s. This is shown in Figure \ref{pvslx}, where
also the maximum observed luminosities are plotted. The X--ray
pulsars are generally divided into different classes, based on
the spectral type classification of the mass donor companion star.
Several excellent reviews describe  the properties
of accretion powered  pulsars (Nagase 1989, White et al. 1995,
Bildsten et al. 1997). Here I will
only comment on a few recent developments  in this field.

\section{\saxj : The Missing Link}

Most  X--ray pulsars have massive companions, either OB
supergiants or Be stars. On the other hand the more numerous
X--ray sources in Low Mass binary systems, display a variety of
interesting phenomena indicating the presence of  an accreting
neutron star, but are characterized by the lack of periodic
coherent signals identifiable as the neutron star spin. Numerous
searches for periodicities have been carried out in Low Mass
X-ray Binaries (LMXRB) without success for more than 15 years, until the
recent discovery of pulsations at 2.5 ms in the transient source
\saxj (Wijnands \& van der Klis 1998).

This finding is of extreme importance, since the  LMXRB
are thought to be the progenitors of the
millisecond radio pulsars (see, e.g., Bhattacharya 1995).
It is believed
that the weakly magnetized neutron stars in LMXRB are spun-up to
rotation periods of a few millisecond due to their interaction
with the rapidly rotating  inner part of the accretion  disk.
Once accretion terminates, the X-ray emission stops and the
neutron star can shine again in the radio band  as a \textit{recycled} rotation powered
pulsar.
The detection of a very rapidly rotating neutron star in \saxj finally
provided a clear evidence supporting
this evolutionary scenario.

However, this finding also raises the puzzling question of why
coherent pulsations have been seen    only in one system, out of
several tens of objects in which a similar signal could have been
detected. It is possible that this is due to an orientation effect
that makes the periodic modulation visible in \saxj . In fact,
pulse arrival time measurements in this source yielded an orbital period
of $\sim$2 hours, a projected semimajor axis of
63 light-ms  and a very small mass function of $\sim$3.8$\times$10$^{-5}$
$\msole$ (Chakrabarty \& Morgan 1988).
This indicates that \saxj has   a  light companion ($<$ 0.2 $\msole$)
and probably a very low inclination ($<$20$^{\circ}$).

Interestingly, \saxj is a transient system, i.e. a source
that spends most of the time in  a state of low luminosity.
Although the mechanism responsible for the mass accretion
variations that cause the transient behavior of these systems is
still unclear,  transient X--ray binaries give the possibility to
study the physics of the accretion and of the interaction of the
matter with the neutron star magnetosphere over a wide interval
of accretion rates (Campana et al. 1998).
A recent observation, carried out with the
BeppoSAX satellite, has provided a measurement of the quiescent
emission from \saxj at a  luminosity  level of only
$\sim$2-3$\times$10$^{32}$ erg s$^{-1}$ (Stella et al. 2000).
This luminosity is too high to be due to the coronal emission from
the companion star. Different possibilities to explain
the  observed X--ray flux have been considered, but the observational
data, at the limit of the BeppoSAX capabilities,
did not provide enough statistics for a spectral study
that could discriminate among the different models.
It is possible that, as observed in other  soft X--ray
transients,
the  quiescent spectrum of \saxj contain also a
power law tail, in addition to the softer component
due to the thermal emission from the neutron star atmosphere.
Stella et al. (2000) discussed the constraints that can be derived
on two possible models for the power law spectral component:
accretion stopped at the magnetospheric radius
or shock emission from the interaction between the relativistic
wind of the neutron star and the wind from the companion.

Detailed spectral and timing studies of \saxj
will be possible with the new, more sensitive
X-ray satellites, such as XMM Newton.
Thus, future observations of the ``missing link'' \saxj,
exploiting the knowledge    of a well determined
spin period (so far not available for other
soft X--ray transients), will greatly  help to understand the nature of the
quiescent emission from X--ray transient sources.

\section{X-ray Pulsars in the Magellanic Clouds}

With an angular size of a few square degrees,
the  Magellanic Clouds are ideal targets for imaging X--ray telescopes.
They offer the advantage of providing a population of sources
at the same, known distance
(respectively $\sim$54 and $\sim$60 kpc, for the Large and Small Magellanic
Cloud).

The current list of all the known Magellanic Clouds X--ray pulsars  is
reported in Table 1.
Most of the  X--ray pulsars in the Magellanic Clouds
have   been discovered only in the last few years,
thanks to observations with the ASCA and
BeppoSAX satellites.

It seems that such a  large
number of newly discovered sources  cannot simply be due to the
fact that these  satellites have
devoted much more time to observe our satellite galaxies than
previous X--ray missions.
Especially in the Small Cloud, the pulsars in neutron star/Be systems
seem to be  more abundant than in our
 Galaxy.
The  total mass of the Small Magellanic Cloud is  about 100 times
smaller than that of our Galaxy.
Scaling the observed  number of SMC Be pulsars  by this factor we should see $\sim$2000
such objects in our Galaxy, compared to  the $\sim$40 actually observed.

In reality the discrepancy is not as large as these numbers
would suggest, due to the presence of selection
effects that make more difficult to observe X-ray sources in our Galaxy.
In fact, the galactic Be/neutron star pulsars
have a  rather
flat distribution in galactic longitude. This indicates that their
average distance is   only  a few kiloparsecs, i.e. we are not sampling
the whole galactic volume: we can only see the closest systems.
Another selection effect is related to the transient nature of the majority
of these systems. Dim transients in the galactic plane are difficult
to discover, due to the limited coverage with sensitive instrument
(all sky monitors have relatively high flux thresholds and are often
confusion limited at low galactic latitudes) and to  the effect
of interstellar absorption.

However, although these selection effects are difficult to quantify,
they are probably not large enough to
completely explain the   overabundance of massive
X--ray binaries in the Small Magellanic Cloud, that is probably
related to the different star formation history in our satellite galaxies.
In fact another related evidence is the paucity of Low Mass X--ray binaries
in the Magellanic
Clouds. Despite all the recent observations only a few bright
LMXB  are known, in striking
difference with the corresponding galactic population.

\begin{table}[]
\centerline{\bf Table 1 -  X--ray Pulsars in the Magellanic Clouds}
\vspace{0.5cm} 
\label{mctab}
\begin{tabular}{|l|c|l|l|}

\hline
\multicolumn{4}{|c|}{ {\bf  Pulsars in the Large Magellanic Cloud } }                      \\
\hline
{\bf  NAME}       &{\bf P (s)}  &{\bf Class$^{(a)}$}  &{\bf References}         \\
\hline
        A 0538--66          &         0.069&      Be      &      Skinner et al. 1982                    \\
        RX J0502.9--6626    &          4.06&      Be      &      Schmidtke et al. 1995                   \\
        LMC X-4            &          13.5&      S       &      Kelley et al. 1983, Vrtilek et al. 1997         \\
        EXO 053109--6609.2  &         13.67&      Be?     &      Dennerl et al. 1996                 \\
        RX J0529.8--6556    &          69.5&      Be      &      Haberl et al. 1997                 \\
        SAX J0544.1--710    &        96.08 &      Be      &      Cusumano et al. 1998                   \\
\hline
\multicolumn{4}{|c|}{ {\bf  Pulsars in the Small Magellanic Cloud } }                      \\

\hline
{\bf  NAME}       &{\bf P (s)}  &{\bf Class$^{(a)}$}  &{\bf References}         \\
\hline
        AX J0043--737       &      0.0876  &      ?       &      Yokogawa \& Koyama 2000      \\
        SMC X-1            &         0.717&      S       &      Lucke et al 1976        \\
        SMC X-2            &        2.37  &      Be      &      Corbet \& Marshall 2000                 \\
        RX J0059.2--7138    &      2.763   &      Be      &      Hughes 1994          \\
        AX J0105--722       &         3.34 &      Be?     &      Yokogawa \& Koyama 1998      \\
        XTE J0052--723      &       4.782  &      Be?     &      Corbet et al. 2001 \\
        2E 0050.1--7247     &           8.9&      Be      &      Israel et al. 1997       \\
        AX J0049--732       &         9.13 &      Be?     &      Imanishi et al. 1998         \\
        RX J0052.1--7319    &          15.3&      Be      &      Lamb et al. 1999       \\
        RX J0117.6--7330    &         22.07&      Be      &      Macomb et al. 1999         \\
        XTE J0111.2--7317   &         30.95&      Be      &      Yokogawa  et al. 2000a       \\
        1WGA J0053.8--7226  &         46.63&      Be      &      Corbet et al. 1998          \\
        1SAX J0054.9--7226  &            59&      Be      &      Marshall \& Lochner 1998         \\
        RX J0049.1--7250    &        74.67 &      Be      &      Yokogawa et al. 1999            \\
        AX J0051--722       &         91.12&      Be      &      Corbet et al. 1998          \\
        AX J0057.4--7325    &        101.42&      Be?     &      Torii et al. 2000           \\
        XTE J0054--720      &           169&      Be?     &      Lochner et al. 1998          \\
        AX J0051.6--7311    &         172.4&      Be      &      Yokogawa et al. 2000b         \\
        AX J0058--7203      &         280.3&      Be?     &      Tsujimoto et al. 1999         \\
        AX J0051--733       &         323.2&      Be      &      Imanishi et al. 1999            \\
        2E 0101.5--7225     &         343.5&      Be      &      Israel et al. 2000           \\
        AX J0049.5--7323    &         755.5&      Be      &      Yokogawa et al. 2000c           \\
\hline
\end{tabular}
\vspace{0.5cm} 

Notes:
(a) S = Supergiant companion; Be = Companion of Be spectral type;  A question mark indicates
that the optical counterpart has not yet been identified and the source is classified only
on the basis of its X--ray properties.
\centerline{\bf~~}

\end{table}

\section {The Anomalous X-ray Pulsars}

In the last few years it has been recognized
(Mereghetti \& Stella 1995, van Paradijs et al. 1995) that
there is a   class of X--ray pulsars with  properties clearly
different from those of the more common   pulsars accreting from
high mass companions.
These objects have been called   Anomalous X--ray Pulsars (AXP,
see Mereghetti (2000) for a detailed review of their properties).
Six AXP are currently known, three of which are clearly
associated with Supernova Remnants (see Table 2).

AXP have  spin periods in a narrow range ($\sim$6-12 s), compared
to the much broader one (0.069 - $\sim$10$^4$ s) observed in
HMXRB pulsars (see Fig.~1). Their periods are monotonically
increasing,   on timescales of $\sim$10$^4$ - 4$\times$10$^5$
years, again at variance with the typical behavior of the
majority of  accreting pulsars that are either spinning-up or
display an erratic period evolution. While their P and $\pdot$ values
strongly suggest that  AXP  are  neutron stars, it is clear that
the corresponding rotational energy loss ($\sim$10$^{45}$
$\Omega$ $\Omdot$ erg s$^{-1}$) is not sufficient to power the
luminosities    of these objects, that are typically in the range
10$^{34}$ - 10$^{36}$ erg s$^{-1}$.

The optical counterparts of AXP are not known
(with the possible exception of \uu, see below).
On the basis of
the limits in the optical and IR bands, the presence of a massive
companion star (OB  super giants and/or Be stars) can be excluded
in most AXP. Furthermore,  there are no signatures of orbital
motion in their X-ray light curves (i.e. no periodic
modulations/eclipses nor Doppler shifts in the spin frequency
induced by an orbital motion of the source).

The AXP are characterized by soft X-ray spectra, clearly
different from those of the pulsars in HMXRB. The latter have
relatively hard spectra in the 2-10 keV range (power law photon
index $\alpha_{ph}\sim$1) that steepen with an exponential
cut-off above $\sim$20 keV. Observations with the ASCA and
BeppoSAX satellites, have shown that in most cases a single power
law is not sufficient to describe the spectra of AXP. All the AXP
for which good quality observations are available (White et al.
1996, Parmar et al. 1998, Oosterbroek et al.1998, Israel et al.
1999a) require the combination of a blackbody-like component with
kT$\sim$0.5 keV, accounting up to  $\sim$40-50\% of the observed
luminosity, and a steep ($\alpha_{ph}\sim$3--4)  power law (see
Table~2). The emitting area inferred from the blackbody
components (R$_{BB}\sim$ 1-4 km) corresponds to a large fraction
of the neutron star surface.

\begin{table}[]
\label{axp}
\centerline{\bf Table 2 - The Anomalous X--ray Pulsars}
\vspace{0.5cm} 
\begin{tabular}{|c|c|c|c|c|}

\hline

{\bf  SOURCE}       &{\bf P (s)}  &{\bf $\pdot$  (s s$^{-1}$)}  &{\bf SNR}       &  {\bf    SPECTRUM  }  \\
                   &             &                    & {\bf d (kpc)/age (kyr)}  &  {\bf kT$_{BB}$/$\alpha_{ph}$}    \\
\hline
1E~1048.1--5937    & 6.45 &[1.5--4]$\times$10$^{-11}$ &    --                 & BB+PL  [3] \\
                   & [1]  &      [2,3]                &                       & $\sim$0.64 keV / $\sim$2.5\\
1E~2259+586        & 6.98 & $\sim$5$\times$10$^{-13}$ & G109.1--0.1 [7,8,9]   & BB+PL [9]                    \\
                   & [4]  &       [5,6]               &   4--5.6 / 3--20      & $\sim$0.44 keV / $\sim$3.9\\
4U~0142+61         & 8.69 & $\sim$2$\times$10$^{-12}$ &    --                 & BB+PL [11,12]                \\
                   & [10] &       [11]                &                       & $\sim$0.4 keV /  $\sim$4   \\
RXSJ170849--4009   &11.00 & 2$\times$10$^{-11}$       &    --                 & BB+PL [13]                \\
                   & [13] &       [14]                &                       & $\sim$0.41 keV/ 2.92  \\
1E~1841--045       &11.77 & 4.1$\times$10$^{-11}$     &  Kes 73 [17,18]       & PL  [19]                \\
                   & [15] &     [16]                  &   6--7.5 / $\lsim$3   &  -- / $\sim$3.4  \\
AX~J1845.0--0300   & 6.97 & --                        & G29.6+0.1 [21]        & BB  [20]    \\
                   & [20] &                           &     $<$20 / $<$8      & $\sim$0.7 keV / --  \\

\hline
\end{tabular}
\vspace{0.5cm} 

[1] Seward et al. 1986;
[2] Mereghetti 1995;
[3] Oosterbroek et al. 1998;
[4] Fahlman \& Gregory 1981;
[5] Baykal \& Swank 1996;
[6] Kaspi et al. 1999;
[7] Hughes et al. 1984;
[8] Rho \& Petre 1997;
[9] Parmar et al. 1998;
[10] Israel et al. 1994;
[11] Israel et al. 1999a;
[12] White et al. 1996;
[13] Sugizaki et al. 1997;
[14] Israel et al. 1999b;
[15] Vasisht \& Gotthelf 1997;
[16] Gotthelf et al. 1999;
[17] Sanbonmatsu \& Helfand 1992;
[18] Helfand et al. 1994;
[19] Gotthelf \& Vasisht 1997;
[20] Torii et al. 1998;
[21] Gaensler et al. 1999;
\centerline{\bf~~}

\end{table}

Though the absence of a massive companion and the presence of a
neutron star are  well established, the AXP remain one of the
more enigmatic classes of galactic X--ray sources:
the   mechanism responsible for the observed X--ray luminosity is
still  unclear. As mentioned above,    models powered by the
rotational energy loss of isolated neutron stars can be excluded
on energetic grounds. The models based on neutron stars proposed
for the AXP involve either  accretion (with or without a binary
companion of very low mass) or the  decay of a very strong
magnetic field. Also more exotic possibilities involving quark
stars have been discussed (Dar \& De Rujula 2000).

\subsection{Accretion based models for the AXP }

Binary models have the advantage of naturally providing accretion
as a  source of energy. However, the  tight limits on the
possible companion stars   have also   led to
interpretations based on accretion on isolated neutron stars.

Mereghetti \& Stella (1995) originally proposed that the AXP are
weakly magnetized neutron stars (B$\sim$10$^{11}$ G) rotating
close to their equilibrium  period. This requires accretion rates
of the order of a few 10$^{15}$ g s$^{-1}$, consistent with the
AXP luminosities.

The possible nature of the companion stars
is constrained directly by the optical/IR limits on the AXP  counterparts
and indirectly by the absence of orbital Doppler modulations of the pulses.
The first method allows to exclude bright massive companions, while the
limits on the Doppler modulations are now beginning to exclude also main sequence stars for
large regions of the orbital parameter space.
Except for the   unlikely possibility that these systems are seen face-on,
main sequence companions can be ruled out
in the three best studied AXP (\ee, \oo, \uu;
Mereghetti, Israel \& Stella 1998, Wilson et al. 1998).
 Helium burning stars
with M $\lsim 0.8\msole$   cannot be excluded,
but the accretion rate resulting from
Roche lobe overflow would produce a much greater luminosity than
the observed one.
A possibility is that the He companion underfill its Roche lobe thus giving a
smaller accretion rate  by a stellar wind.
White dwarf companion stars are compatible with the a$_x\sin i$
limits and yield consistent values of accretion. For example,
a white dwarf of   $\sim0.02~\msole$
and  P$_{orb}\sim$   30 min would give the
$\mdot$ of a few $\times 10^{-11}~\msole$~yr$^{-1}$ required by the
observed luminosity of 1E~2259+586.

Accretion from the interstellar medium  (ISM)
cannot provide the luminosities observed in AXP for typical
ISM parameters and neutron star velocities.
The accretion luminosity is given by
L$_{acc}\sim$10$^{32}$~~v$_{50}^{-3}$~~n$_{100}$~erg s$^{-1}$,
where  v$_{50}$ is the relative velocity between the
neutron star and the ISM in units of 50 km s$^{-1}$ and
n$_{100}$ is the gas density in units of 100 atoms cm$^{-3}$.
Unless all the AXP lie within nearby ($\sim$100 pc) molecular clouds,
which seems very unlikely considered their distribution
in the galactic plane, the accretion rate   is
clearly insufficient to produce the observed luminosities.

An alternative possibility involving isolated  neutron stars fed
from a residual accretion disk was first advanced by Corbet et
al. (1995) for \ee , and developed in more detail by van Paradijs
et al. (1995) and Ghosh et al. (1997). These authors proposed that
AXP   result from the common envelope
evolution    of close massive X-ray binary systems. The connection with
massive binaries is supported by the fact that the AXP seem to be
relatively young objects, being located at small distances from
the galactic plane and, in at least 50\% of the cases,
at the center of SNRs. A residual accretion disk could be formed after the
complete spiral-in of a neutron star in the envelope of a giant
companion (a Thorne-Zytkow object, TZO, Thorne \& Zytkow 1977).
According to
Ghosh et al. (1997), a massive binary undergoing
common envelope evolution can produce two kinds of objects,
depending on the (poorly known) efficiency with which the
envelope of the massive star is lost. Relatively wide systems
have enough orbital energy to lead to the complete expulsion of
the envelope before the settling of the neutron star at the
center of the massive companion. This results in the formation of
binaries composed of a neutron star and a helium star, like
4U~1626--67 and Cyg X--3. Closer HMXRB, on the other hand,
produce TZO, due to the complete spiral in of the neutron star in
the common envelope phase, and then evolve into AXP.

According to Ghosh et al. (1997), this model can also explain
the two component X--ray spectra of AXP, as well as  their secular
spin-down.
The accretion flow is supposed to consist of two distinct components:
one  forming   a  disk and one spherically symmetric,
resulting from the part of the envelope with
less angular momentum.
The hot (kT$\sim$1 keV) and ionized
spherically symmetric flow forms a shock
at the magnetospheric boundary, cools, and
enters into the magnetosphere through a  Rayleigh-Taylor
instability. This results in accretion over
 a large fraction of the neutron star surface, producing the
observed blackbody emission.
The power law spectral component is instead produced
by the conventional, field-aligned accretion onto the polar caps
resulting   from the  disk component.
The  AXP are supposed to rotate close to their equilibrium
periods, which increase  due to the decreasing mass accretion rate
(see, however, Li (1999) for a criticism to this model).

Another possibility for the formation of a disk around an isolated
neutron star is through fallback of some material from the
progenitor star after the supernova explosion
(Chatterjee et al.  2000).
For appropriate values of the neutron star magnetic field,
initial spin period, and mass of the residual disk,
these systems can evolve into AXP with   luminosities, periods and lifetimes
consistent with the observed values.
Due to the steadily declining mass accretion rate,
the rotating neutron star evolves through different states.
During an initial ``propeller'' phase, lasting a few thousand years, the
spin period increases up to values close to those observed in AXP.
In this phase, the AXP progenitors are  very faint, undetectable X--ray sources,
since accretion down to the neutron star surface is inhibited
(or greatly reduced) by the magnetospheric centrifugal barrier.
In the following phase, the spin frequency approaches the Keplerian
frequency at the inner edge of the disk $\Omega$(r$_m$), most of the mass flow is
accreted, and the star becomes visible as an AXP. During this
quasi-equilibrium phase, the neutron star  spins down trying to
match  $\Omega$(r$_m$), which decreases with the declining mass accretion rate in the disk.
To explain  the narrow range of spin periods observed in AXP,
Chatterjee et al. (2000)
propose that an advection-dominated accretion flow (ADAF)
ensues when the accretion rate further decreases. This causes a
very small X--ray luminosity, thus explaining the
lack of old  AXP ($\gsim$5 10$^4$ yrs) with long
spin periods.

\subsection{Strongly magnetized neutron stars (Magnetars)}

Models based on
strongly magnetized (B$\sim$10$^{14}$--10$^{15}$ G) neutron stars,
or \textit{``magnetars''} (Duncan \& Thompson 1992; Thompson \& Duncan 1995, 1996),
were originally developed to explain
the Soft Gamma-ray Repeaters (SGR).
SGRs are remarkable transient events
characterized by brief ($<$ 1 s) and relatively soft (peak photon energy
$\sim$20-30 keV) bursts of super-Eddington luminosity.
Only four (or possibly five) SGRs are currently known
(see Hurley 2000 for a review).
Several authors
pointed out some analogies between the prototype AXP \ee and
the soft repeater  SGR 0526--66, located in the Large Magellanic Cloud
SNR N49 and for which pulsations at 8 s were reported during
the famous super-burst of March 5, 1979.

The possible connection between AXP and SGR,
received renewed attention
after the discovery of periodicities
also in SGR 1806--20  (Kouveliotou et al. 1998)
and  SGR~1900+14 (Hurley et al. 1999, Kouveliotou et al. 1999).
The values of P and $\pdot$
($\sim$ (5-15) 10$^{-11}$ s s$^{-1}$)
observed in SGRs are very similar
to those of AXP. Other similarities with the AXP are the luminosity
of their quiescent counterparts,
(L$_X$ $\sim$10$^{34}$-10$^{35}$ erg s$^{-1}$)
and the fact that all of them appear to be associated with SNRs.

If the spin-down in  AXP and SGR is interpreted as due
to magnetic dipole radiation losses, the
neutron star magnetic field can be estimated as
B $\sim$ 3.2$\times$10$^{19}$ ($P\pdot$)$^{1/2}$ G.
The observed values of P and $\pdot$ lead to values
of B$\gsim$10$^{14}$-10$^{15}$ G.
In the magnetar model the magnetic field
is the main energy source, powering both the  persistent
X--ray (and particle) emission  and the soft gamma-ray bursting
activity. This  involves  internal heating, due to the
magnetic field dissipation,  and the
generation of seismic activity. The latter is responsible
for the soft $\gamma$-ray  bursts, when the magnetic stresses in the
neutron star crust shake the magnetosphere
and accelerate particles.

Heyl \& Hernquist (1997) showed that,
if the  magnetic fields in AXP are  $\gsim$10$^{15}$ G, their residual thermal energy
can be  sufficient to power for a few thousand years
the observed X--ray luminosity.
This requires the presence of an   envelope of
hydrogen and helium (an iron envelope is much more
efficient in insulating  the core, resulting in a
lower luminosity and effective temperature at the
neutron star surface). The envelope of light
elements, with a required mass of $\sim$10$^{-11}$-10$^{-8}$ $\msole$,
could be due to fallback material after the supernova
explosion and/or to accretion from the interstellar medium
if the neutron star is born in a sufficiently dense environment
($\gsim$10$^4$ cm$^{-3}$).

Actually, the AXP magnetic fields derived through the dipole
radiation formula are very likely  overestimated.
In fact, the particle wind outflow,
either continuous or in the form of strong
episodic outbursts,  also contributes significantly to the spin-down
(Thompson \& Blaes 1998).   Harding et al. (2000)  estimated magnetic field and spin-down age
as a function of the particle wind duty cycle and luminosity.
They found that,
in the case of  continuous particle  outflows, the AXP and SGR
magnetic fields can be in the same range as those
of conventional radio pulsars.

Colpi et al. (2000) noted that, in the context of the magnetar
scenario, the period clustering of AXP can be explained only if
the magnetic field decays on a timescale of $\sim$10$^4$ years.
Models without a significant field decay would lead to
the presence of AXP with longer periods, which have not been observed.

\subsection{Discriminating between AXP models}

Different authors discussed the kind of  spin-down irregularities
expected in  the magnetar model.
Melatos (1999) described the
oscillation in $\pdot$  caused by radiative precession,
an effect due to the star asphericity induced by the
very strong magnetic field. He fitted the observed
evolution of rotation frequency
of the AXP \ee and \oo
in terms of the  periodic ($\sim$5-10 yrs)
behavior of $\pdot$ resulting from this effect.
Unfortunately,  the sparse period measurements available
for AXP do not allow, for the moment, to discriminate
against alternative possibilities.

For instance,  Heyl \& Hernquist (1999)
fitted the same data with a constant spin-down interrupted
by a few   glitches.  The   magnitude of these glitches
is similar to that observed in radio pulsars.
Earlier analysis of the same \ee data   showed that the level of $\pdot$ fluctuations
was similar to that of accreting
X--ray pulsars
(Baykal \& Swank 1996), and was therefore taken as support to
accretion based models.
More recently, Kaspi et al. (1999) could
obtain for \ee a phase-coherent timing solution, thanks to RXTE observations
spanning 2.6 years. These data show
a very low level of timing noise, contrary to the
previous results     based on sparse
observations that could not be phase related.
Also \rx , monitored with RXTE for 1.4 yrs, was found to have a low
level of timing noise
(Kaspi et al. 1999),
until the detection of a sudden spin-up event with all the characeristics of
a glitch (Kaspi et al. 2000).

In conclusion it seems that timing studies can certainly help to understand
AXP, but the emerging picture is still unclear. The low level of timing
noise observed in the pulsars mentioned above (and in \kes, Gotthelf
et al. 1999) contrasts with the more irregular behavior of \oo
(Paul et al. 2000).

The other promising way to  discriminate between different AXP models
is through deep optical observations. Recently   Hulleman et al. (2000)
reported the discovery of a faint (R$\sim$25) blue object  in the error box
of \uu. According to these authors, the faintness of the proposed optical
counterpart rules out the presence of an accretion disk, thus favoring
the magnetar interpretation. Unfortunately, detailed estimates
of the expected optical brightness from disks around isolated neutron
stars are very uncertain and depends on several factors, like
the disk inclination, dimensions, amount of X--ray reprocessing, etc...
(Perna et al. 2000).
It seems therefore premature to draw firm conclusions based on the
single case of \uu.
The search for the optical/IR counterparts of    other AXP is complicated
by the large reddening and by the fact that their error boxes are not
small enough to search for counterparts at such faint magnitude levels
(see, e.g. the case of \kes, Mereghetti et al. (2001)).
In this respect, more
accurate positions for the AXP are expected from  the on-going program
of observations with the Chandra
and XMM Newton satellites, which will also provide high quality spectral
information, possibly allowing to discriminate between different X--ray
emission mechanisms.

%

%
%
\section { References}





\reff Baykal A.  \&   Swank J.H. 1996, ApJ 460, 470.

\reff Becker W. 2000, Advances in  Space Research 25, 647.


\reff Bhattacharya D. 1995, in {\it X--ray Binaries}, eds. W.H.G. Lewin, J. van Paradijs \& E.P.J. van den Heuvel, (Cambridge Cambridge Univ. Press, 233.

\reff Bildsten L. et al. 1997, ApJS 113, 367.

\reff Campana S., Colpi M., Mereghetti S., Stella L., Tavani M. 1998, A\&AR 8, 269.

\reff Chakrabarty D. \& Morgan E.H. 1998, Nature 394, 346.

\reff Chakrabarty D. et al.   1998, IAU Circ. 7048.  

\reff Chatterjee P., Hernquist L., \& Narayan R. 2000, ApJ 534, 373.   


\reff Colpi M., Geppert U., \& Page D. 2000, ApJ 529, L29. 

\reff Corbet R.H.D, Smale A.P., Ozaki M. et al. 1995, ApJ  443, 786. 

\reff Corbet R.,  Marshall F.E. 2000, IAU Circ 7402.  

\reff Corbet R. et al. 1998, IAU Circ. 6803.  

\reff Corbet R. et al. 2001, IAU Circ. 7562.

\reff Cusumano G.,  Israel G.L.,  Mannucci F.,  Masetti N.,  Mineo T.,  Nicastro L.    1998, A\&A 337,    772. 


\reff Dar A. \& De Rujula A. 2000, astro-ph/0002104

\reff Davidson K. \& Ostriker J. 1973, ApJ 179, 585.


\reff Dennerl K.,  Haberl F.,  Pietsch W.   1996, Proc. WuMPE Report 263,    131. 

\reff Dubus G., et al. 1999, MNRAS 302, 731.

\reff Duncan R.C. \& Thompson C.  1992, ApJ 392,  L9. 

\reff Fahlman G.G. \& Gregory, P.C. 1981, Nature 293, 202. 


\reff Gaensler B.M., Gotthelf E.V. \& Vasisht G. 1999,  ApJ 526, L37.  


\reff Ghosh P., Angelini L. \& White N.E. 1997, ApJ 478,  713.

\reff Gotthelf E.V. \& Vasisht G. 1997, ApJ 486, L133.


\reff Gotthelf E.V., Vasisht G. \& Dotani 1999, ApJ 522, L49. 


\reff Haberl F.,  Dennerl K.,  Pietsch W.,  Reinsch K.     1997, A\&A 318,    490. 

\reff Harding A.K., Contopoulos I., \&  Kazanas D. 2000, ApJ 525, L125. 

\reff Helfand D. et al. 1994, ApJ 434, 627. 


\reff Heyl J.S. \& Hernquist L.  1997, ApJ  489, L67.

\reff Heyl J.S. \& Hernquist L.  1999, MNRAS 304, L37. 

\reff Hughes V.A. et al. 1984, ApJ 283, 147. 

\reff Hughes J.P.     1994, ApJ 427,    L25. 

\reff Hulleman F., van Kerkwijk M.H. \& Kulkarni S.R., 2000, Nature, 408, 689.


\reff Hurley K. et al.  1999, ApJ 510, L111. 

\reff Hurley K. 2000, Proceedings 5th Huntsville GRB Symposium, AIP 526, 763.

\reff Imanishi K. et al. 1998, IAU Circ. 7040.  

\reff Imanishi K.,  Yokogawa J.,  Tsujimoto M.,  Koyama K. 1999, PASJ 51,    L15. 

\reff Israel G.L., Mereghetti, S., Stella L. 1994, ApJ 433, L25. 

\reff Israel G.L., et al. 1995, IAU Circ. 6156.

\reff Israel G.L.,  Stella L.,  Angelini L.,  White N.E.,  Giommi P.,  Covino S. 1997, ApJ 484,    L141. 


\reff Israel G.L., et al. 1999a, A\&A 346, 929. 

\reff Israel G.L., et al. 1999b, ApJ 518, L107.  

\reff Israel G.L.,  Campana S.,  Covino S.,  Dal Fiume D.,  Gaetz T.J.,  Mereghetti S.,  Oosterbroek T.   2000, ApJ 531,    L131. 


\reff Kaspi V.M.,  Chakrabarty D. \&  Steinberger J. 1999, ApJ 525, L33. 

\reff Kaspi V.M., Lackey J.R. \& Chakrabarty D.   2000, ApJ 537, L31.  

\reff Kelley R.L.,  Jernigan J.G.,  Levine A.,  Petro L.D.,   Rappaport S. 1983, ApJ 264,  568. 


\reff Kouveliotou C. et al. 1998, Nature  393, 235. 


\reff Kouveliotou C. et al. 1999, ApJ 510, L115. 


\reff Lamb R.C. et al. 1999, IAU Circ. 7081.  

\reff Li X.-D. 1999, ApJ 520, 271. 

\reff Lochner J.C. et al. 1998, IAU Circ. 6814.  

\reff Lucke R.,  Yentis D.,  Friedman H.,  Fritz G.,  Shulman S.  1976, ApJ 206,    L25. 

\reff Macomb D.J.,  Finger M.H.,  Harmon B.A.,  Lamb R.C.,  Prince T.A. 1999, ApJ 518,    L99. 

\reff Marshall F.E.,  Lochner J.C.   1998, IAU Circ. 6818. 

\reff Melatos A. 1999, ApJ 519, L77. 

\reff Mereghetti S. 1995, ApJ 455, 598. 

\reff Mereghetti S. \& Stella L. 1995, ApJ  442, L17.


\reff Mereghetti S., Israel G.L.  \&  Stella L. 1998, MNRAS  296,  689. 

\reff Mereghetti S. 2000, Proceedings  NATO ASI ''The Neutron Star - Black Hole Connection'', in press, astro-ph/9911252

\reff Mereghetti S. et al. 2001, MNRAS, 321, 143.


\reff Nagase F. 1989, PASJ 41, 1.

\reff Oosterbroek T., Parmar A.N., Mereghetti S. \& Israel G.L. 1998,  A\&A 334, 925.


\reff Parmar A. et al. 1998,  A\&A 330, 175. 

\reff Paul B., Kawasaki M., Dotani T. \& Nagase F. 2000, ApJ 537, 319. 

\reff Perna R., Hernquist L. \& Narayan R. 2000, ApJ 541, 344. 


\reff Pringle J.E. \& Rees M.J. 1972, A\&A  21, 1.


\reff Rho J. \& Petre R.  1997, ApJ  484, 828. 


\reff Sanbonmatsu K.Y. \& Helfand D.J. 1992, AJ 104, 2189. 


\reff Schmidtke P.C.,  Cowley A.P.,  McGrath T.K.,  Anderson A.L. 1995, PASP 107,    450. 

\reff Seward F., Charles, P.A., Smale, A.P. 1986, ApJ 305, 814. 

\reff Skinner G.K.,  Bedford D.K.,  Elsner R.F.,  Leahy D.,  Weisskopf M.C.,  Grindlay J. 1982, Nature 297, 568. 


\reff Stella L., Campana S., Mereghetti S., Ricci D., Israel G.L. 2000, ApJ 537, L11.

\reff Sugizaki M. et al. 1997, PASJ 49, L25. 


\reff Thompson C. \& Duncan R.C. 1995, MNRAS 275, 255. 

\reff Thompson C. \& Duncan R.C. 1996, ApJ  473,  322.


\reff Thompson, C. \& Blaes O. 1998, Phys. Rev. D 57, 3219.

\reff Thorne K.S. \& Zytkow A.N. 1977, ApJ   212, 832.

\reff Torii K. et al. 1998, ApJ   503,  843. 

\reff Torii K. et al. 2000, IAU Circ. 7441. 

\reff Tsujimoto M.,  Imanishi K.,  Yokogawa J.,  Koyama K. 1999, PASJ 51,    L21. 


\reff van Paradijs J., Taam R.E. \& van den Heuvel E.P.J. 1995, A\&A 299, L41.


\reff Vasisht G.  \&   Gotthelf E.V. 1997, ApJ 486, L129. 


\reff Vrtilek S.D.,  Boroson B.,  Cheng F.H.,  McCray R.,  Nagase F.  1997, ApJ 490,    377. 


\reff White N.E. et al. 1996, ApJ 463, L83. 

\reff White N.E., Nagase, F, \& Parmar, A.N. 1995, in ``X--ray Binaries'',  eds. W.H.G. Lewin, J. van Paradijs \& E.P.J. van den Heuvel  (Cambridge Cambridge Univ. Press), 1.


\reff Wijnands R. \& van der Klis M. 1998, Nature 394, 344.

\reff Wilson C.A., Dieters S., Finger M.H., Scott D.M. \& van Paradijs J. 1998, ApJ 513, 464. 


\reff Yokogawa J.,  Koyama K.  1998, IAU Circ. 7028. 

\reff Yokogawa J.,  Imanishi K.,  Tsujimoto M.,  Kohno M.,   Koyama K.  1999, PASJ 51,    547. 

\reff Yokogawa J.,  Koyama K.  2000, IAU Circ. 7361.  

\reff Yokogawa J., Paul B., Ozaki M., Nagase F., Chakrabarty D., Takeshima T. 2000a, ApJ 539, 191.

\reff Yokogawa J.,  Torii K.,  Imanishi K.,  Koyama K. 2000b, PASJ 52, L37.  

\reff Yokogawa J.,  Imanishi K., Ueno M., Koyama K. 2000c, PASJ, in press

\end{document}